# How can AI reduce wrist injuries in the workplace?

R.F. Pitzalis[1,2], N. Cartocci[1,3], C. Di Natali[1], D. G. Caldwell[1], G. Berselli[1,2], J. Ortiz[1]
[1]XoLab, Advanced Robotics, Istituto Italiano di Tecnologia (IIT), Genoa, Italy.
[2]Department of Mechanical, Energy and Transportation Engineering (DIME), University of Genoa, 16145, Genoa, Italy.
[3]Department of Informatics, Bioengineering, Robotics and Systems Engineering (DIBRIS), University of Genoa (UniGe), Genoa, Italy.

Speaker: R.F. Pitzalis

## Abstract

This paper explores the development of a control and sensor strategy for an industrial wearable wrist exoskeleton by classifying and predicting workers' actions. The study evaluates the correlation between exerted force and effort intensity, along with sensor strategy optimization, for designing purposes. Using data from six healthy subjects in a manufacturing plant, this paper presents EMG-based models for wrist motion classification and force prediction. Wrist motion recognition is achieved through a pattern recognition algorithm developed with surface EMG data from an 8-channel EMG sensor (Myo Armband); while a force regression model uses wrist and hand force measurements from a commercial handheld dynamometer (Vernier GoDirect Hand Dynamometer). This control strategy forms the foundation for a streamlined exoskeleton architecture designed for industrial applications, focusing on simplicity, reduced costs, and minimal sensor use while ensuring reliable and effective assistance.

## 1. Introduction

Among all types of exoskeletons, in recent decades, very few have focused on the wrist, especially for industrial applications, as reported in many review articles on upper limb exoskeletons [1 - 5], as most are developed for rehabilitation and training post-injury and have limitations in design and control. Control strategies using multi-sensor data—such as position (via encoders, IMUs), force/torque (using load cells, FSRs), and bio-signals (EMG)— have increased over the years and provide critical feedback from the user to drive exoskeleton actuators. Traditional control methods, such as Conventional Trajectory Tracking (CTT) and Assist-As-Needed (AAN), rely on predefined dynamic models, while newer machine learning (ML) and deep learning (DL) approaches bypass complex equations by predicting human movement patterns from measured data [6 - 10]. In wrist exoskeletons, a common ML control strategy involves classifying surface electromyography (sEMG) signals to detect user intention and control actuators accordingly [6 - 10]. However, only a few studies address activity recognition and force estimation together. Leone et al. [11] introduced a hierarchical model based on Non-Linear Logistic Regression (NLR) for gesture and force level classification, while other studies [12] and [13] focus on predicting grip force and wrist angles using simultaneously sEMG and accelerometers. This research seeks to enhance sEMG-based control for wrist exoskeletons by accurately predicting user actions and efforts while optimizing EMG sensor placement. By identifying the minimal EMG channels required, we aim to balance accuracy with computational efficiency for force prediction across wrist activities. This approach guides the design of an industrial wrist exoskeleton and helps determine if additional sensors are necessary for effective assistance. The paper is organized as follows: Section II covers system architecture and data acquisition; Section III details EMG and force data processing, feature extraction, and reduction; Section IV discusses algorithms for gesture classification and force estimation; and Sections V and VI summarize main achievements and outline future work.

## 2. Methodology

This chapter briefly describes the aims of this research and focuses on the experimental method of acquiring EMG and force measurements developed from wrist movements.

### 2.1 System design

This research is intended as a preliminary study for designing and developing a prototype of a soft wrist exoskeleton to assist primary wrist movements, provide sufficient force support, and ensure user acceptance in industrial settings. The exoskeleton will actuate and monitor three degrees of freedom—wrist flexion/extension, radial/ulnar deviation, and grasping—using sEMG sensors to detect muscle activity and predict wrist movement and exerted force. A two-level control strategy will be used, with high-level control interpreting EMG signals to guide movement and provide assistive force, and low-level control ensuring precise tracking at each actuator to guarantee that the user can move freely as intended and experience assistive forces with appropriate timing and intensity. This paper outlines a high-level control strategy using machine learning to classify wrist movements and forces exerted from raw EMG data. The classified movement determines the direction of the force/torque and, thus, which motor to activate based on the expected movement, while EMG signal amplitude determines the output force/torque intensity to be provided by the actuators.

### 2.2 Experimental setup

To detect and predict wrist movements, forearm muscle activity is recorded with a Myo Armband (Thalmic Labs), an 8-channel, dry-electrode EMG sensor with a 200 Hz sampling rate. While more convenient than gel-based electrodes, it has limitations in signal quality, sensitivity to motion artifacts, and frequency range (recommended EMG sampling is 1000 Hz for EMG signals with a frequency range of 5-500 Hz) [9], [10]. The acquisition system further down-samples the signals to 100 Hz, which may reduce the accuracy of ML classifiers in distinguishing wrist gestures. Despite these limitations, the Myo Armband is chosen for its ease of use, even in production settings. Additionally, a GoDirect Hand Dynamometer (by Vernier) measures wrist force across various positions. Data is collected on a Raspberry Pi 4, with the Myo Armband streaming wirelessly to a USB dongle on the Pi, and the dynamometer connected via USB. At the same time, the execution and recording program is run from a host computer terminal. Figure 1 shows the setup.

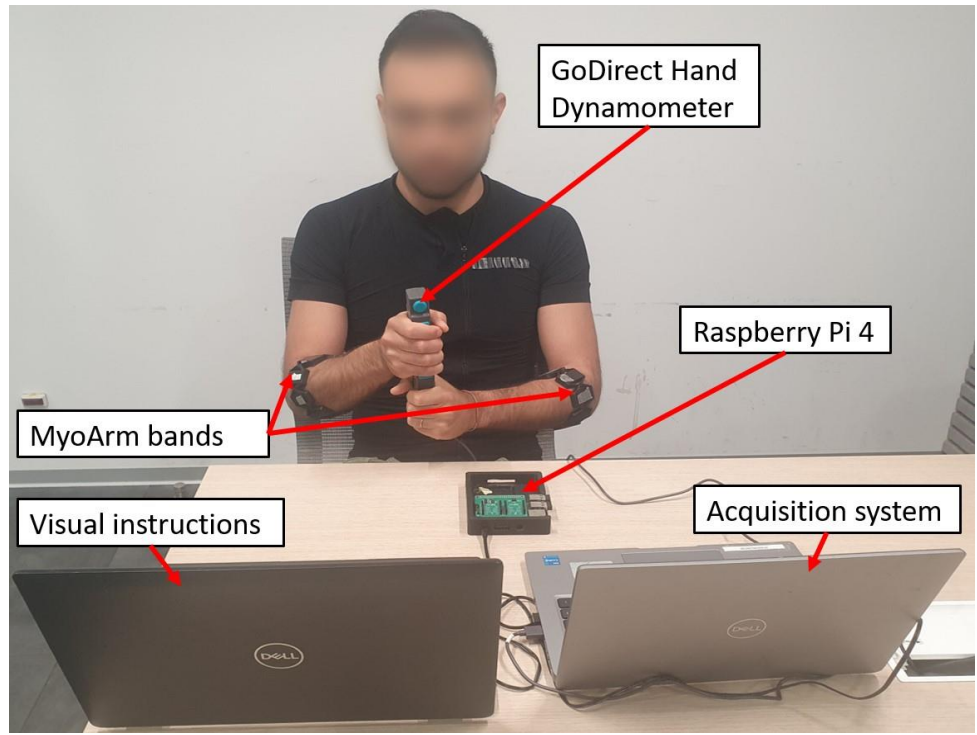

*Figure 1 - Setup of the data acquisition system. Example of measurement during a Hand Close (i.e., grasping) phase.*

## 2.3 Experimental protocol

Participants in this study are artisan workers from Prestige Italia S.p.A., a leading horse saddle manufacturer. Their work is physically demanding, often leading to injuries such as carpal tunnel syndrome. Six healthy male subjects (ages 25–45) were recruited, five right-handed and one left-handed. The Liguria Ethics Committee approved the experiment (CER Liguria 001/2019), adhering to the Helsinki Declaration and GDPR for privacy protection [14]. For each experimental session, participants were seated with elbows and forearms resting on a table, leaving only the wrist free to move, ensuring minimal shoulder and elbow compensation. Visual instructions guided the participants through five wrist gestures, shown in Figure 2, — Wrist Flexion (WF), Wrist Extension (WE), Radial Deviation (WRD), Ulnar Deviation (WUD), and Hand Close (HC, or grasping) — chosen to align with the exoskeleton's degrees of freedom. Each participant wore a Myo Armband on both forearms to capture EMG signals. They are positioned following the arrangement proposed in [9], and about the orientation of the Myo LED light indicating channel No. 4, the EMG sensors are positioned near the following muscles: *Extensor Carpi Ulnaris* (channel 1), *Extensor Digitorum* (channel 2), *Extensor Carpi Radialis Longus e Brevis* (channel 3), *Brachioradialis* (channel 4), *Flexor Carpi Radialis* (channel 5) *Palmaris longus* (channel 6) *Flexor Digitorum Superficialis* (channel 7), and *Flexor Carpi Ulnaris* (channel 8). To ensure that the same channels read signals from the same left and right muscles, the Myo Armbands are mirrored on the forearms to reflect the anatomical arrangement muscle bundles. Participants performed each movement isometrically to reach the Maximum Voluntary Contraction (MVC) of the forearm muscles while applying maximum force on a digital dynamometer. This helped correlate EMG signal amplitude with exerted force and distinguish muscle activation for each gesture. Data collection for each task followed this sequence: starting with the wrist at rest (5-6s), then moving into the designated position while applying maximum contraction, holding for

5-6s, and finally relaxing for 5-6s. For Hand Close, participants also performed a step-force task, gradually increasing grip from 0N to 100 N by 10N increments every 5-6s. Since grasping is one of the most frequent gestures active during manipulation and is often combined with other wrist movements, this measure evaluates muscle activation across different force levels, especially during sustained activities with force requests far from the maximum (at most 1/3). Participants did not show any signs of physical or mental fatigue.

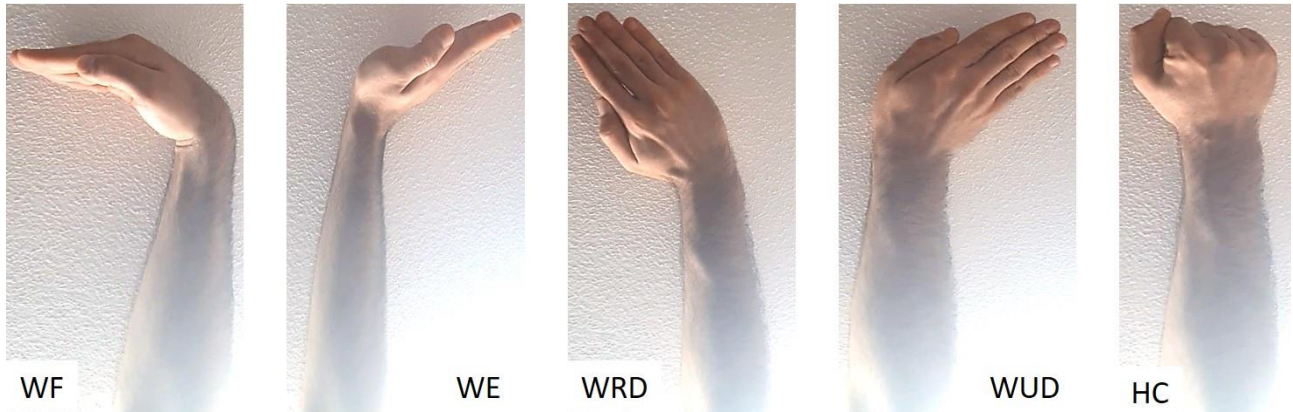

*Figure 2 - Wrist gestures included in the study: from left to right Wrist Flexion (WF), Wrist Extension (WE), Wrist Radial Deviation (WRD), Wrist Ulnar Deviation (WUD), and Hand Close (HC).*

## 3. Data Processing

This chapter presents the processing of EMG data, as schematically illustrated in Figure 3, with a description of the various techniques used to extract features and statistical indices from the dataset for classification.

### 3.1 EMG Pre-processing

The electromyographic signal is inherently noisy, so it undergoes pre-treatment before use. First, signals are rectified by converting data to absolute values, then passed through a 5 Hz low-pass filter optimized for this dataset. Finally, they are normalized to the 95th percentile of Maximum Voluntary Contraction (MVC) to ensure comparability, smooth out electrical spikes, and produce a cleaner signal.

### 3.2 Channel selection

For Gesture Recognition (GR) and Force Estimation (FE), a data-driven feature selection approach is used with EMG data. The Minimum Redundancy Maximum Relevance (mRMR) [15] algorithm is applied to select the most informative and distinct channels. Widely used in fields like cancer diagnostics and speech recognition [16], [17], mRMR minimizes redundancy in the feature set while maximizing relevance to the target variable by analyzing mutual information between features and the response variable. or each feature, a score is calculated to assess its importance. Each feature's score is normalized for the total sum, and the most relevant channels are chosen for GR and FE.

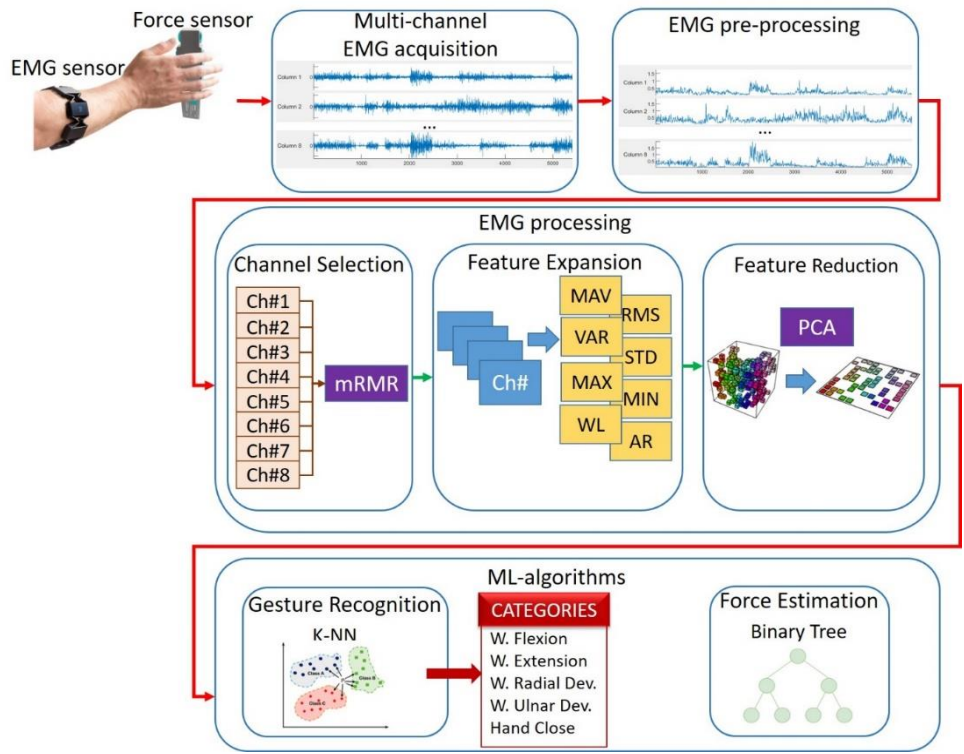

*Figure 3 - Block diagram describing the data acquisition process, the EMG signal classification, and the force prediction scheme.*

## 3.3 Features Expansion

Feature expansion enhances the dataset by generating new features based on the original set, creating a subspace with additional statistical indices from the Time Domain (variation of each EMG signal over a time window) and Spatial Domain (variation across EMG channels at the same sampling time). Time-based features are widely used in EMG recognition due to their ease of calculation [18]. The following features are calculated (summarized in Table I, with N as the time window and K as the number of channels):

- Mean Absolute Value (MAV): Average absolute EMG signal amplitude.
- Root Mean Square (RMS): Average signal power over time.
- Variance (VAR): Mean square deviation of the signal.
- Standard Deviation (STD): Square root of variance, indicating data spread.
- Maximum (MAX) and Minimum (MIN): Highest and lowest signal values.
- Waveform-Length (WL): Cumulative signal length, reflecting its waveform amplitude, frequency, and duration.
- Auto-Regressive (AR) Model: Expresses each sample as a linear combination of previous samples, using AR model coefficients as features.

As the number of channels changes, feature expansion adjusts the feature set accordingly.

## 3.4 Features Reduction

After feature expansion, Principal Component Analysis (PCA) reduces data dimensions by projecting the dataset onto a lower-dimensional subspace. This technique creates uncorrelated variables that retain the most relevant information and capture maximum variance from the original data. The top transformed features are selected based on cumulative variance. This reduces model complexity, minimizes over-fitting, and enhances computational efficiency and predictive accuracy [19]. The number of components after PCA analysis is shown in Table II. Unlike feature expansion, the number of PCA features can vary depending on explained variance and data distribution, even when the number of channels remains fixed.

## 4. Results

The control algorithm described is based on state-of-the-art controllers that utilize machine-learning strategies to predict hand/wrist movements according to muscle activation and force [9 - 11], [14], [15]. Instead of more advanced neural networks or deep learning techniques, machine learning has been preferred and considered more efficient due to limited experimental data [19].

### 4.1  Gesture Recognition (GR)

The Gesture Recognition (GR) part estimates wrist movement using the 8 EMG channels of the Myo Armband. This classification problem is solved with the K-Nearest Neighbors (KNN) algorithm [20], chosen over Support Vector Machine (SVM) and neural network due to comparable or superior results with less computational demand. Neural networks performed poorly given the need for more subjects and repetitions, so alternative techniques were not further detailed. After data preprocessing, the mRMR algorithm assesses the importance of each channel in the Myo Armband, showing that channels 2, 5, and 8 are the most significant. Selected channels undergo feature expansion, and then reduction with PCA to retain 95% explained variance. Using these features, a KNN model is trained on data from six subjects performing five wrist movements (WF, WE, WUD, WRD, and HC) for a total of 60 sequences and approximately ~45k samples, employing 5-fold cross-validation. Model parameters include 10 neighbors, Euclidean distance, and standardized data. Accuracy increases with the number of channels: averaging 50% with one channel, 70% with two, 85% with three, and over 90% with more. Figure 4 illustrates this accuracy trend. In Figure 5, GR results for the fourth subject compare an 8-channel model (99% accuracy) with a 3-channel model (87% accuracy) against ground truth. While 8 channels provide higher accuracy, the 3-channel model achieves adequate performance with reduced computational load for real-time applications.

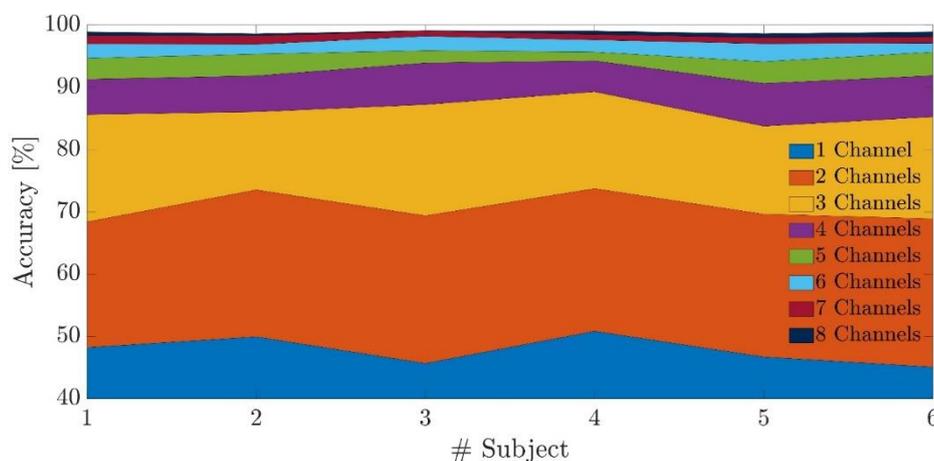

*Figure 4 - Accuracy obtained from KNN models as a function of the number of selected channels (from 1 to 8 channels as displayed in the legend) for different subjects.*

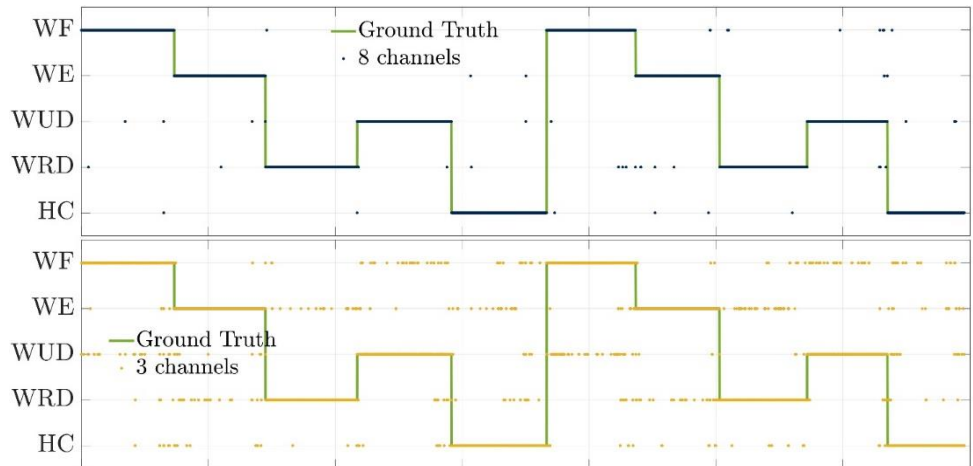

*Figure 5 - Comparison of Gesture Classification accuracy for the KNN model with eight channels (top) and three channels (bottom) available for different tasks of the fourth subject. The green line describes the ground truth, while the dark blue/yellow dots represent the expected value of the gesture.*

## 4.2   Force Estimation (FE)

The Force Estimation (FE) process predicts force output using the 8 EMG signals from the Myo Armband. This problem is addressed with a Regression Tree algorithm, chosen over neural networks due to limitations in subject and repetition variety. After data preprocessing, the mRMR algorithm ranks the importance of each EMG channel, with channel 8 rated highest, followed by channels 2 and 5. Selected channels undergo feature expansion and PCA reduction, retaining 95% of the variance. Using these features, a Regression Tree model is trained on data from six subjects performing five wrist activities for both hands, yielding 72 sequences and ~110k samples using a 5-fold cross-validation scheme with the following parameters: a minimum leaf size of 10 and a low-pass filter with a 1Hz cutoff. Force values, ranging from a few to hundreds of Newtons, are normalized by subtracting the minimum value and dividing by the 95% percentile to enhance prediction accuracy, with the regressor output filtered at 1Hz to smooth quantized results. Models trained with different numbers of channels are evaluated using Root-Mean-Square Error (RMSE) and Median Absolute Percentage Error (MdAPE). With one channel, the average MdAPE is around 22%, dropping to 13% with two channels, 9% with three, and below 5% with more. Three channels result effective for FE as well as Gesture Recognition. Moreover, the most effective channels are the same already chosen for GR. Figures 6 and 7 compare FE results for all activities in the fourth subject's left and right hands respectively, using an eight-channel (dark blue lines) and three-channel model (yellow) against the ground truth (green). Although the three-channel model is slightly noisier, it remains accurate, supporting efficient real-time force estimation for wrist exoskeleton control.

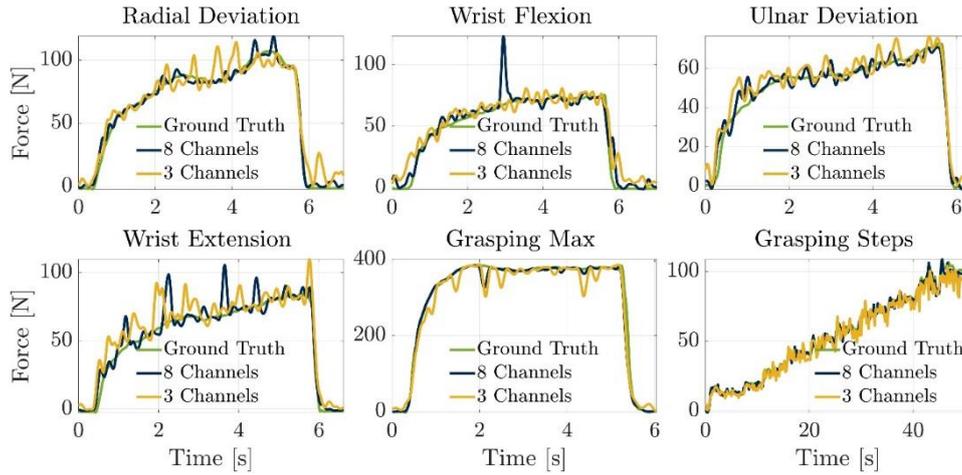

*Figure 6 - Comparison of the Force Estimation for the filtered Regression Tree model with eight channels (dark blue lines) and three channels (yellow lines) for all the activities conducted with the fourth subject's left hand. The green line describes the ground truth.*

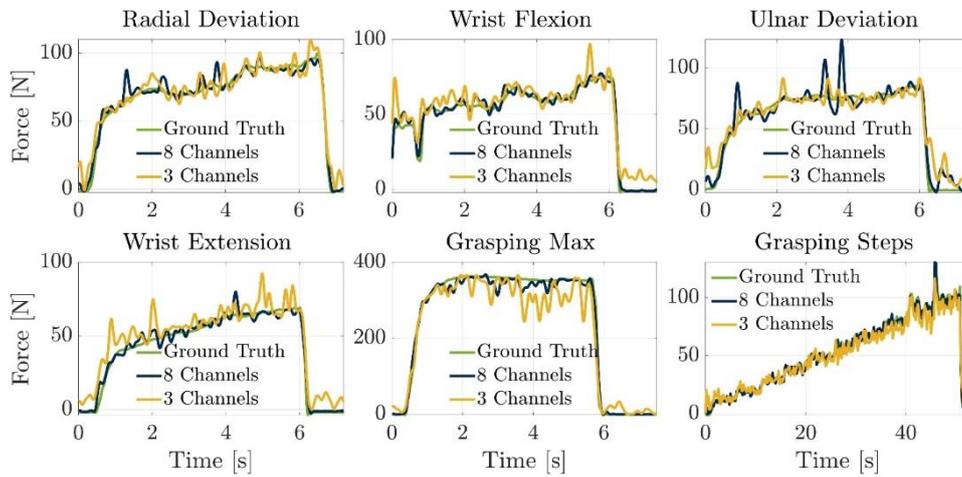

*Figure 7 - Comparison of the Force Estimation for the filtered Regression Tree model with eight channels (dark blue lines) and three channels (yellow lines) for all the activities conducted with the fourth subject's right hand. The green line describes the ground truth.*

## 5. Discussion

Results in Section IV indicate that using eight EMG channels achieves a GR accuracy of 99%, which decreases to 87% when reduced to three channels. For FE, MdAPE is below 5% with eight channels and around 9% with three. Despite these small declines, using three channels is sufficient to get remarkable results, enabling real-time operation with lower computational demand. Notably, the key channels for both GR and FE are channel 2, channel 5, and channel 8 correspond to the *Extensor Digitorum*, *Flexor Carpi Radialis*, and *Flexor Carpi Ulnaris* muscles, respectively. This insight is valuable for wrist exoskeleton design, as positioning three EMG sensors optimally could be sufficient to track wrist motion and estimate force/torque requirements without needing IMU or force sensors. A limitation of this study is the limited amount of data. While a free dataset [9] is partially helpful, the subjects in these studies are researchers with different muscle activation and joint development compared to field workers. Therefore, integrating data from such sets with that collected at Prestige Italia poses challenges.

## 6. Conclusion

Wrist exoskeletons require advancements in design and control strategies for improved reliability and precision. Modeling wrist kinematics and dynamics are challenging due to their complexity, making machine learning (ML) valuable for learning, classifying, and predicting user-specific movement patterns without predefined equations. Our research develops a control strategy for an industrial wrist exoskeleton using ML for pattern recognition to identify wrist movements (flexion, extension, radial and ulnar deviations, grasping) and a regression model to estimate the force exerted. An EMG sensor strategy optimization is also considered and evaluated to simplify the design, costs, and data process by adopting as few sensors and electronic components as possible while retaining accuracy and reliability and evaluating whether additional sensors are necessary. The findings show that three EMG electrodes positioned on key forearm muscles (Extensor Digitorum, Flexor Carpi Radialis, Flexor Carpi Ulnaris) can achieve over 85% accuracy in predicting wrist movement and estimating force. Data from six workers are collected using an 8-channel EMG sensor (Myo Armband) and a hand dynamometer (GoDirect by Vernier). After data pre-processing (filtering, normalization), feature selection (mRMR), feature expansion (statistical indices MAV, RMS, VAR, etc.), and PCA-based reduction, significant features are identified for different channels. A Gesture Recognition (GR) algorithm via KNN and a Force Estimation (FE) model via a Regression Tree predict wrist motions and force output. This approach supports both hands and lays a foundation for advanced control in a wrist exoskeleton for industrial assistance. Limitations include the small sample size and repetitions; future work aims to improve accuracy with a larger dataset, training the model to assist users in various tasks and adapt to individual differences.